\documentclass[preprint,12pt]{elsarticle}
\usepackage[utf8]{inputenc}
\usepackage{amssymb,graphicx,subfigure,amsmath,epsfig,multicol,multirow,dsfont,bm,enumerate,upgreek,hhline,color,url,upgreek,stmaryrd,nicefrac}
\usepackage{rotating} 

\journal{NeuroImage: Clinical}
\begin{document}
\begin{frontmatter}

\title{White matter hyperintensity and stroke lesion segmentation and differentiation using convolutional neural networks}
\author[icl]{R. Guerrero\corref{cor1}\fnref{fn1}}
\ead{reg09@imperial.ac.uk}
\cortext[cor1]{Corresponding author: Ricardo Guerrero}
\fntext[fn1]{These authors contributed equally to this work.}
\author[icl]{C. Qin\fnref{fn1}}
\author[icl]{O. Oktay}
\author[icl]{C. Bowles}
\author[icl]{L. Chen}
\author[ixico]{R. Joules}
\author[ixico,icl]{R. Wolz}
\author[edi]{M.C. Vald\'{e}s-Hern\'{a}ndez}
\author[edi]{D.A. Dickie}
\author[edi]{J. Wardlaw}
\author[icl]{D. Rueckert}

\address[icl]{Department of Computing, Imperial College London, UK}
\address[ixico]{IXICO plc., UK}
\address[edi]{UK Dementia Research Institute at The University of Edinburgh, Edinburgh Medical School, 47 Little France Crescent, Edinburgh, EH16 4TJ, UK}

\begin{abstract}
White matter hyperintensities (WMH) are a feature of sporadic small vessel disease also frequently observed in magnetic resonance images (MRI) of healthy elderly subjects. The accurate assessment of WMH burden is of crucial importance for epidemiological studies to determine association between WMHs, cognitive and clinical data; their causes, and the effects of new treatments in randomized trials. The manual delineation of WMHs is a very tedious, costly and time consuming process, that needs to be carried out by an expert annotator (e.g. a trained image analyst or radiologist). The problem of WMH delineation is further complicated by the fact that other pathological features (i.e. stroke lesions) often also appear as hyperintense regions. Recently, several automated methods aiming to tackle the challenges of WMH segmentation have been proposed. Most of these methods have been specifically developed to segment WMH in MRI but cannot differentiate between WMHs and strokes. Other methods, capable of distinguishing between different pathologies in brain MRI, are not designed with simultaneous WMH and stroke segmentation in mind. Therefore, a task specific, reliable, fully automated method that can segment and differentiate between these two pathological manifestations on MRI has not yet been fully identified. In this work we propose to use a convolutional neural network (CNN) that is able to segment hyperintensities and differentiate between WMHs and stroke lesions. 
Specifically, we aim to distinguish between WMH pathologies from those caused by stroke lesions due to either cortical, large or small subcortical infarcts. 
The proposed fully convolutional CNN architecture, called uResNet, is comprised of an analysis path, that gradually learns low and high level features, followed by a synthesis path, that gradually combines and up-samples the low and high level features into a class likelihood semantic segmentation. Quantitatively, the proposed CNN architecture is shown to outperform other well established and state-of-the-art algorithms in terms of overlap with manual expert annotations. Clinically, the extracted WMH volumes were found to correlate better with the Fazekas visual rating score than competing methods or the expert-annotated volumes. Additionally, a comparison of the associations found between clinical risk-factors and the WMH volumes generated by the proposed method, were found to be in line with the associations found with the expert-annotated volumes.
\end{abstract}


\end{frontmatter}

\section{Introduction}
\subsection{Clinical motivation}
White matter hyperintensities (WMH), referred to in the clinical literature as leukoaraiosis, white matter lesions or white matter disease \cite{wardlaw13}, are a characteristic of small vessel disease \cite{wardlaw14} commonly observed in elderly subjects on fluid-attenuated inversion recovery (FLAIR) magnetic resonance (MR) images, which, as the name suggests, they appear as hyperintense regions. 
Moreover, stroke lesions of cortical, large subcortical (striatocapsular) or small subcortical infarct origin can also often appear as hyperintense regions in FLAIR MR images and can coexist and coalesce with WMHs. 
The accurate assessment of WMH burden 
is
of crucial importance for epidemiological studies to determine associations between WMHs, cognitive and clinical data. Similarly, it would help discover their causes, and the effects of new treatments in randomized trials.
In the assessment of WMH 
burden it is important to exclude stroke lesions as they have different underlying pathologies, and failure to account for this may have important implications for the design and sample size calculations of observational studies and randomized trials using WMH quantitative measures, WMH progression
or brain atrophy as outcome measures \cite{wang12}.
One of the most widely used metrics to assess WMH burden and severity is the Fazekas visual rating scale (i.e. score) \cite{fazekas87}. In this scale, a radiologist visually rates deep white matter and peri-ventricular areas of a MR scan into four possible categories each depending on the size, location and confluence of lesions. The combination of both deep white matter and peri-ventricular ratings yields a combined zero to six scale. In the vast majority of clinical trials and in general clinical practice visual rating scores are used (such as the Fazekas score).
WMHs are very variable in size, appearance and location, and therefore
the categorical nature of the Fazekas scale has limitations for studying their progression in relation with other clinical parameters. WMH volume has been demonstrated to correlate with severity of symptoms, progression of disability and clinical outcome \cite{bendfeldt10,chard02,louvbld97}. Accordingly, determining WMH volume has been of interest in clinical research as well as in clinical trials on disease-modifying drugs \cite{louvbld97,van-gijn98,brott89,spirit97}. For some studies, lesions have been traced manually (sometimes with the help of semi-automated tools for contour detection) slice by slice. This process can easily become prohibitively expensive for even moderately large datasets.
It is therefore obvious that the accurate automatic quantification of WMH volume would be highly desirable, as this will undoubtedly lead to savings in both time and cost.
Recently, several automated and semi-automated methods have been put forward to address the coarseness of the visual assessments (e.g. Fazekas score), as well as the dependence on highly qualified experts to perform such assessments. These methods can be broadly classified into \textit{supervised}, when a training ``gold-standard'' is available \cite{van_nguyen15,ghafoorian16}, i.e. when one or more human experts have annotated data, \textit{unsupervised}, when no such gold-standard exists \cite{ye13,cardoso15,bowles16}, and \textit{semi-supervised}, when only a small portion of available data has been expertly annotated \cite{kawata10,qin16}. However, despite the number of proposed methods, no automated solution is currently widely used in clinical practice and only a few of them are publicly available \cite{shiee10,damangir12,schmidt12}.  
This is partly because lesion load, as defined in most previously proposed automatic WMH segmentation algorithms, does not take into account the contribution of strokes lesion, as these methods are generally unable to differentiate between these two types of lesions. 

\subsection{Related work}
In the following we review existing methods and challenges that are related to our work, especially on Multiple sclerosis (MS), WMH and stroke lesion segmentation in MR imaging. 
Additionally, some more general CNN segmentation approaches that share architectural similarities with the method we propose here are also reviewed in this section. 
Over the last few years, there has been an increased amount of research going on in these areas \cite{garcia2013review,caligiuri2015automatic,maier2017isles,rekik2012medical}. Although some of the methods mentioned here were proposed for segmenting different pathologies rather than the ones we explore in this work, they can in fact be applied to different tasks. As mentioned before, these methods can be broadly classified into \textit{unsupervised}, \textit{semi-automatic}, \textit{semi-supervised} and \textit{supervised}, depending on the amount of expertly annotated data available.

\paragraph{Unsupervised segmentation}
Unsupervised segmentation methods do not require labeled data to perform the segmentation. Most of these approaches employ clustering methods based on intensity information or some anatomical knowledge to group similar voxels into clusters, such as fuzzy C-means methods \cite{gibson2010automatic}, EM-based algorithms \cite{dugas2004improved,forbes2010adaptive,kikinis1999quantitative} and Gaussian mixture models \cite{freifeld2009multiple,khayati2008fully}. Some of the probabilistic generative models of the lesion formation for stroke lesion segmentation were also designed, such as  \cite{forbes2010adaptive,derntl2015stroke}.
Forbes et al. \cite{forbes2010adaptive} proposed a Bayesian multi-sequence Markov model for fusing multiple MR sequences to robustly and accurately segment brain lesions. 
Derntl et al. \cite{derntl2015stroke} proposed to combine standard atlas-based segmentation with a stroke lesion occurrence atlas, in a patient-specific iterative procedure.
Some authors have also proposed to model lesions as outliers to normal tissues. Van Leemput et al. \cite{van2001automated} employed a weighted EM framework in which voxels far from the model were weighted less in the estimation and considered potential lesions. Weiss et al. \cite{weiss13} proposed to use dictionary learning to learn a sparse representation from pathology free brain T1-weighted MR scans and then applied this dictionary to sparsely reconstruct brain MR images that contain pathologies, where the lesions were identified using the reconstruction error. Additionally, 
several works have also focused on exploiting the fact that WMHs are best observed in FLAIR MR images, while being difficult to identify in T1-weighted MR images. Some of these methods rely on generating a synthetic FLAIR image based on observed T1-weighted MR image using random forests \cite{ye13}, generative mixture-models \cite{cardoso15}, support vector regression (SVR) \cite{bowles16} or convolutional neural networks (CNN) \cite{van_nguyen15}. Both synthetic (healthy looking) and real FLAIR (with pathologies) images are then compared to detect any abnormalities. 
Other method like lesion-TOADS \cite{shiee2010topology} combines atlas segmentation with statistical intensity modeling to simultaneously segments major brain structures as well as lesions.
The lesion growth algorithm (LGA), proposed by Schmidt et al. \cite{schmidt12} and  part of SPM's LST toolbox (www.statistical-modelling.de/lst.html), constructs a conservative lesion belief map with a pre-chosen threshold ($\kappa$), followed by the initial map being grown along voxels that appear hyperintense in the FLAIR image. In essence, LGA is a self-seeded algorithm and it tends to have difficulties detecting subtle WMHs. An important drawback of all these methods is that they are in fact abnormality detection algorithms and not specifically WMH segmentation methods, hence in principle they detect any pathology, whether or not is a WMH-related pathology.

\paragraph{Semi-automatic and semi-supervised segmentation}
Several semi-automatic algorithms proposed in the literature for WMH segmentation rely on region growing techniques that require initial seed points to be placed by an operator. Kawata et al. \cite{kawata10} introduced a region growing method for adaptive selection of segmentation by using a SVM with image features extracted from initially identified WMH candidates. Itti et al. \cite{itti01} proposed another region growing algorithm that extracts WMHs by propagating seed points into neighboring voxels whose intensity is above an optimized threshold. The process iterates until convergence, i.e. all voxels above the threshold that are connected to the initial seed point had been annotated. Aside from the drawback of requiring per image expert inputs, semi-automatic methods have the additional potential drawback that seeds points could easily be selected in obvious regions, while the biggest challenge of WMH segmentation can arguably be found in the more confusing border regions. Qin et al. \cite{qin16} proposed a semi-supervised algorithm that optimizes a kernel based max-margin objective function which aims to maximize the margin averaged over inliers and outliers while exploiting a limited amount of available labeled data. Although theoretically interesting and well motivated, the problem of transferring useful knowledge from unlabeled data to a task defined by partially annotated data remains a challenge and an open field of research in its own right. Hence, in practice, semi-supervised WMH segmentation methods, even though they still require some expert input, tend to underperform when compared to supervised methods, even when the later are trained with only a modest amount of data.

\paragraph{Supervised segmentation}
Supervised methods for lesion segmentation have also been well researched. Classical supervised machine learning methods such as k-nearest neighbors (kNN) \cite{anbeek2004probabilistic}, Bayesian models \cite{maillard2008automated}, support vector machines (SVM) \cite{lao2008computer}, and random forests \cite{geremia2011spatial} have been well studied in MS segmentation. For stroke lesion segmentation, pattern classification techniques to learn a segmentation function were also employed in \cite{prakash2006identification,maier2014ischemic,maier2015extra,maier2015classifiers}.
The lesion prediction algorithm (LPA) \cite{schmidt2017bayesian}, implemented in SPM's LST toolbox, has been shown to produce consistently good performance and in many cases is considered a robust gold standard for this problem. LPA is described as a logistic regression model, where binary lesion maps of 53 MS patients were used as response values. Additionally, as covariates to this model a lesion belief map similar to those from LGA \cite{schmidt12} was used in combination with a spatial covariate that takes into account voxel specific changes in lesion probability. Recently, Ithapu et al. \cite{ithapu14} proposed using SVMs and random forests in combination with texture features engineered by texton filter banks for WMH segmentation task. Brain intensity abnormality classification algorithm (BIANCA) \cite{griffanti2016bianca}, a fully automated supervised method based on kNN algorithm, was also proposed for WMH segmentation.
An interesting work proposed by Dalca et al. \cite{dalca2014segmentation} used a generative probabilistic model for the differential segmentation of leukoaraiosis and stroke by learning the spatial
distribution and intensity profile of each pathology, which shares the same application purpose with the work proposed here.

More recently, CNNs have been put forward to replace the inference step in many computer vision related tasks \cite{girshick14,long15,he16,dong16}, with current state-of-the-art methods in many fields being dominated by CNN frameworks. CNNs have been shown to have enough capacity to model complex nonlinear functions capable of performing multi-class classification tasks such as those required for the description and understanding of highly heterogeneous problems, such as brain lesion segmentation \cite{brosch2015deep,birenbaum2016longitudinal,kamnitsas15,kamnitsas17, valverde2016multiple,mckinley2016nabla}. For instance, Brosch et al. \cite{brosch2015deep} proposed a deep convolutional encoder network which combines feature extraction and segmentation prediction on MS lesions. Their work was later extended to a 3D deep encoder network with shortcut connections, which consistently outperformed other methods across a wide range of lesion sizes \cite{brosch2016deep}. 
Kamnitsas et al. \cite{kamnitsas17} proposed a network architecture with two parallel convolutional pathways that processes the 3D patches at different scales  followed by a 3D densely connected conditional random field (CRF). Although the method was originally proposed for ischemic stroke, 
tumor and brain injury segmentation on MR images, it can be easily adapted for different tasks using their provided package DeepMedic\footnote{https://github.com/Kamnitsask/deepmedic}.
Similarly, Ghafoorian et al. \cite{ghafoorian16} proposed a CNN architecture that considered multi-scale patches and explicit location features while training, and later was extended to consider non-uniform patch sampling \cite{ghafoorian16b}. Their best performing architecture shares a similar design with the architecture proposed by Kamnitsas et al. \cite{kamnitsas15,kamnitsas17}, in which it trained independent paths of convolutional layers for each scale.

Using multi-resolution inputs \cite{kamnitsas17,ghafoorian16,ghafoorian16b} can increase the field of view with smaller feature maps, while also allowing more non-linearities (more layers) to be used at higher resolution, both of which are desired properties. However, down-sampling patches has the drawback that valuable information is being discarded before any processing is done, and 
since filters learned by the first few layers of CNNs tend to be basic feature detectors, e.g. lines or curves, different paths risk capturing redundant information.
Furthermore, although convolutions performed in 3D as in \cite{kamnitsas17} intuitively make sense for 3D volumetric images, FLAIR image acquisitions are actually often acquired as 2D images with large slice thickness and then stacked into a 3D volume. Further to this, gold standard annotations, such as those generated by trained radiologists (e.g. WMH delineation or Fazekas scores) are usually derived by assessing images slice by slice. Thus, as pointed out by Ghafoorian et al. \cite{ghafoorian16}, 3D convolutions for FLAIR MR image segmentation are in fact less intuitive.

Some other works on CNN segmentation which are relevant to our work, though not on brain lesion segmentation, include Long et al. \cite{long15} and Ronneberger et al. \cite{ronneberger15a}. Long et al. \cite{long15} proposed to segment natural images using a fully convolutional network that supplemented the output of a gradually contracting network with features from several of its levels of contraction through up-sampling. Similar to \cite{long15}, Ronneberger et al. \cite{ronneberger15a} used a U-shaped architecture (U-net) to segment microscopical cell images. The architecture symmetrically combined a contracting and expanding path via feature concatenations, in which up-sampling operations were realized with trainable kernels (deconvolution or transposed convolution). Both of these networks form the foundation of the architecture later proposed in this work.

\paragraph{Challenges}
There are several challenges being held on brain lesion segmentation in recent years. 
For instance, the MS lesion segmentation challenge 2008 (http://www.ia.unc.edu/MSseg/) had the goal of the direct comparison of different 3D MS lesion segmentation techniques. Data used in this challenge consisted of 54 brain MR images from a wide range of patients and pathology severity. 
The 2015 Longitudinal MS Lesion Segmentation Challenge (http://iacl.ece.jhu.edu/index.php/MSChallenge) aimed to apply automatic lesion segmentation algorithms to MR neuroimaging data acquired at multiple time points from MS patients. The ischemic stroke lesion segmentation (ISLES) challenge (http://www.isles-challenge.org/) has been held since 2015, which aims to provide a platform for a fair and direct comparison of methods for ischemic stroke lesion segmentation from multi-spectral MRI image, and asked for methods that allow the prediction of lesion outcome based on acute MRI data. More recently, a WMH segmentation challenge (http://wmh.isi.uu.nl/) was held aiming to directly compare methods for the automatic segmentation of WMH of presumed vascular origin, with data used in this challenge acquired from five different scanners from three different vendors in three different hospitals.

\subsection{Contributions}
\begin{figure}[t]
\centering
\includegraphics[width=12cm]{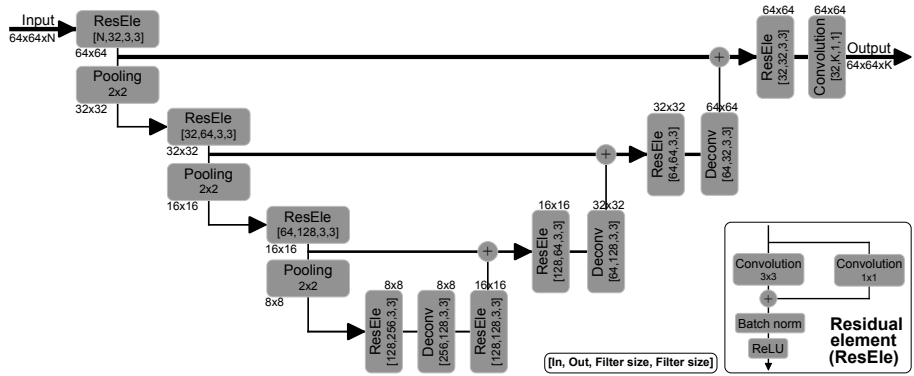} 
\caption{Proposed u-shaped residual network (uResNet) architecture for WMH segmentation and differentiation.}
\label{fig:uResNet}
\end{figure}

In this work we aim to address some of the short comings mentioned before and propose to use a CNN to segment and differentiate between WMH-related pathology and strokes. Specifically, we task ourselves with distinguishing between WMH pathologies from those pathologies originating due to stroke lesions that result from either cortical or subcortical infarcts.
For this, a CNN with an architecture inspired by U-net \cite{ronneberger15a}, originally used to segment neuronal structures in electron microscopic stacks, is proposed. The architecture consists of an analysis path that aims to capture context and a symmetric synthesis path that gradually combines analysis and synthesis features to ultimately enable precise localization. The proposed CNN architecture is trained with large high-resolution image patches and is able to extract high- and low-level features through a single path, thus avoiding filter learning redundancy. Different to \cite{ronneberger15a}, in the work proposed here we replace convolutions with residual elements \cite{he16} and concatenations used in skip connections in the U-net architecture with summations to reduce model complexity. Residual architectures have been shown to ease gradient back-propagation flow, and hence improve optimization convergence speed and allow for deeper network training. An important contribution of this work deals with data sampling for training. Due to the large class imbalance present in WMH segmentation, data sampling for training requires careful consideration, an issue that has received recent research focus due to its influence on the precision of segmentation \cite{kamnitsas17}. Here, to mitigate class imbalance, training is done using patches, rather than dense training on whole images. Further to this, we sample patches that always contain WMH and randomly shift the central location so that WMH can occur anywhere in the patch and not necessarily include the center. As argued before, the proposed CNN architecture is designed for 2D images and it is trained with 2D image patches. Furthermore, we experiment with multi-channel inputs to evaluate the added benefit of adding T1 MR scans and white matter and/or cerebro-spinal track probability maps. The proposed architecture, which we refer as uResNet, can be visualized in Figure \ref{fig:uResNet}.

\section{Methods}
CNNs represent a versatile class of machine learning models that can be trained to predict voxel-wise semantic labels on images. This is achieved by learning a mapping function $f(\Theta, {x}) \rightarrow  {y}$, parametrized by $\Theta$, that transforms voxel level image intensity ${x}$ to a desired label space or image segmentation ${y}\in Y$. Such mapping function $f(\Theta, {x})$ is modeled by a series of $L$ convolution and non-linearity operations, with each element in this sequence generally referred to as a layer. Each layer $l$ produces a set of features maps $H_l$. Here, the convolutional kernel of layer $l$ that produces the $j$th feature map is parametrized as $w_{l}^{j,k}$, where $k$ refers to the $k$th feature map of $H_{l-1}$. The solution to this problem estimates a conditional distribution $p({y}|{x})$ that minimizes the loss function $\Psi$ (see Section \ref{ssec:loss}) defined by ${y}$ and its estimate $f(\Theta, {x})$. After each layer $l$ a set of feature maps or intermediate representations $h_j^{l}$ is obtained. In this work, non-linearities are defined as rectified linear units (ReLU) \cite{nair10}. Intermediate feature maps are computed as convolutions between the convolution kernels $w_{l}^{j,k}$ and the layers' input as
\begin{equation}
 h^j_l=\textup{max} \left( 0, \sum_{k=1}^{J_{l-1}} h^{k}_{(l-1)} * w_{l}^{j,k} \right) \;.
\end{equation}

\noindent Here * denotes the convolution operator, $h^j_0={x}$, and $J_{l-1}$ is the number feature maps in layer $l-1$, with $J_0$ being the number of input channels.

In addition to the sequence of convolution and non-linearity operations mentioned, in the work presented here, residual units or residual elements (ResEle) \cite{he16} are employed to reformulate the previous mapping function as $f(\theta_l, H_{l-1}) + W_l H_{l-1} \rightarrow  H_l$, where $W_l$ performs a linear projection that matches the number of feature maps in layer $l-1$ to those in layer $l$. Figure \ref{fig:uResNet} bottom-right shows the form of ResEle used in this work. Furthermore, to decrease the number of parameters (and control over-fitting) associated with an increase network field-of-view max-pooling layers are employed. Max-pooling operates independently on each input feature map where all but the maximum valued activation with in a support region are discarded, the same is repeated at every strided location. Support region and stride in this work were set to 2x2 and 2, respectively, effectively down-sampling by a factor of two after every max-pool layer.

\subsection{Network architecture}
Defining a  CNN's architecture requires careful consideration of the task is set out to achieve. Important aspects that must to be taken into account are the network's field of view or receptive field and its capacity or complexity. In the architecture proposed here we follow the suggestions of Simonyan and Zisserman \cite{simonyan15} and use only small (3x3) kernels. This allows an increased non-linearity capacity with a lower number of parameters needed for the same receptive field.

The architecture proposed here follows a U-shaped architecture. Furthermore, no fully connected layers are used, thus it is a fully convolutional network, and hence even though it is trained with image patches, inference can be performed on whole images in one single feed forward pass without any need of architectural changes. In total our architecture is composed of 12 layers with $\sim$1M trainable parameters: 8 residual elements, 3 deconvolution layers, and one final convolution layer that converts the feature maps to the label space. Here, the last layer's feature maps $H_L$ are passed to an element-wise softmax function that produces pseudo class probability maps as 
\begin{equation}\label{eq:softmax}
 \rho_c (H_L) = \frac {\textup{exp} (H_L)}{ \sum_{c=1}^{C} \textup{exp} (H_L)} \;\;\; \forall c,
\end{equation}

\noindent where $c$ denotes class and $C$ the total number of classes.

This in essence yields a class-likelihood for each voxel in the image, and its output, in combination with a loss function (described in Section \ref{ssec:loss}), is optimized through the back-propagation algorithm.

\subsection{Loss function and class imbalance}\label{ssec:loss}
In general terms, a loss function maps the values of one or more variables onto a real number that represents some ``cost'' associated with an event. Loss functions defined for classification tasks are functions that calculate a penalty incurred for every incorrect prediction. As mentioned before, casting a semantic segmentation task as a voxel-wise classification problem tends to lead to significant class imbalances. Loss functions can be defined in a such a way that they take class imbalance into account. Here, we will detail a classical loss function that does not take into account class imbalance as well as several recently proposed loss functions that either directly or indirectly take into account class imbalance, as they will be subject of investigation.

In the context of the work presented here, let us define a training set of samples $\textup{X}= \{ \textup{x}_1,...,\textup{x}_{P} \}$, where each $\textup{x}_{p}=\{ {x}_{(p,1)},...,{x}_{(p,V)} \}$ are image patches extracted from in-plane FLAIR (and/or additional modalities) axial slices that will be treated as independent samples during training. Here, $P$ is the total number of patches available and $V$ the total number of voxels per patch. Additionally, let us also define voxel level labels as one-hot encoded variables $y_{p,v}$ associated with each voxel $x_{p,v} \in \textup{X}$. Let us consider $Y\in \mathbb{N}^C$ a one-hot encoded label space, where the class of each voxel in $x_{p,v}$ is given by a $C$-length vector $y_{p,v}$ of all zeros except for a one at position $c$ which indicates the associated label. 
However, let us simplify notation for the following loss equations by re-indexing all voxels in $X$ and their corresponding label as $x_n$ and $y_n$, respectively. Here, $n=\{1,...,N \}$ and $N=P*V$ is the total number of voxels from all patches in $X$.
Therefore, the problem of estimating the mapping function $f(\Theta, x_{n})$ can be defined as the minimization of a loss function that works with the pseudo probabilities obtained from Equation \ref{eq:softmax}. 

A popular loss function for classification tasks, such as the one tackled here, is the categorical cross-entropy which aims to maximize the log likelihood of the data or, equally, minimize the cross-entropy via the following loss function 
\begin{equation}\label{eq:cross-entropy} 
 \Psi=-\sum^{N}_{n=1} y_{n} \textup{log}( f(\Theta, x_{n}) ).
\end{equation}

Classical cross-entropy does not take into account class imbalances in the data which might lead to learning biased predictors. A simple approach to deal with class imbalance that has been proposed for CNN segmentation, is to modify the aggregation of categorical cross-entropy given in Equation \ref{eq:cross-entropy}, by weighting voxels that belong to different classes differently. This modification aims to give more weight to under-represented classes, while weighting down over represented ones, and can be written as
\begin{equation}
 \Psi=-\sum^{N}_{n=1} y_{n} \textup{log}( f(\Theta, x_{n}) )\omega(y_{n}).
\end{equation}
\noindent where $\omega(y_{n})$ is the weight associated to class of $y_{n}$.

Wu et al. \cite{wu16} recently proposed a simple modification of the categorical cross-entropy by dropping, or ignoring, the loss contribution of elements whose correct class prediction was above a certain threshold $\tau$. This has the effect of placing more emphasis on previous mistakes, thus focusing the learning process on ``harder'' (and arguably more valuable) examples during training.
Dubbed online bootstrapped categorical cross-entropy, this loss function can be written as
\begin{equation}
 \Psi=-\sum^{N}_{n=1} y_{n} \textup{log}( \varphi_{n} )
\end{equation}

\noindent where $\varphi_{n} = \{ 1 \;\; \textup{if} \;\; f(\Theta, x_{n})>\tau,\;\; f(\Theta, x_{n}) \;\; \textup{otherwise}\}$.

The Dice coefficient is defined on a binary space and aims at maximizing the overlap between regions of the same class. This makes it a popular and natural choice of metric when comparing binary segmentation labels. However, it is non-differentiable, making its optimization with the back-propagation algorithm not possible. Recently, the winning team of the Second Annual Data Science Bowl\footnote{https://www.kaggle.com/c/second-annual-data-science-bowl}, proposed using a pseudo Dice coefficient as loss function, that can be written as
\begin{equation}\label{eq:dice}
 \Psi=1- \frac{1}{C} \sum^C_{c=1} \left ( \frac{2 \sum^{N}_{n=1} (y^{c}_n f(\Theta,x_n)^c)}{ \sum^{N}_{n=1} f(\Theta,x_n)^c + \sum^{N}_{n=1} y^c_n}\right ).
\end{equation}

\noindent Here, the predicted binary labels are replaced by continuous softmax outputs and averaged across all labels $C$, and where $f(\Theta,x_n)^c$ denotes the softmax prediction of class $c$. Aggregating Dice coefficients from $C$ different classes as an average, has the additional effect of normalizing the per-class loss contribution.

\subsection{Data sampling and class imbalance}\label{sec:sampling}
\begin{figure}[t]
\centering
\includegraphics[width=8cm]{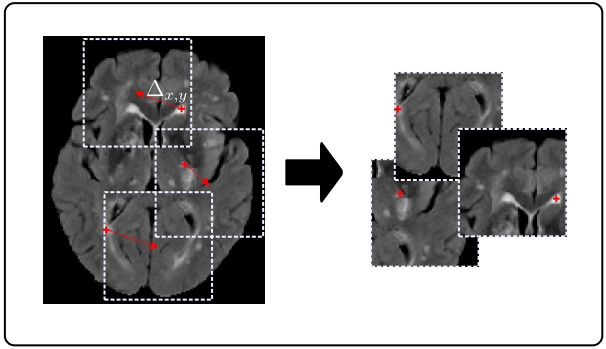} 
\caption{Training patch sampling.}
\label{fig:patch_sampling}
\end{figure}

Generally, in the segmentation of pathologies, healthy tissue is present in far larger quantities than pathological. For example, in WMH segmentation the number of voxels labeled as WMH (regardless of the underlying pathology) is very small compared to those labeled background/healthy tissue, which leads to a significant class imbalance ($\sim$99.8\% of the voxels in the dataset used in this work are labeled as background/healthy tissue in our training set). Hence, although dense training (where whole images or slices are used) is a staple in computer vision with natural images \cite{long15}, it is less intuitive for WMH segmentation. Therefore, patch sampling is used in this work in order to alleviate the class imbalance problem. There are several techniques that could be used to sample patches for training. For example half of the samples could be extracted from locations centered on healthy tissue and half centered on WMH tissue \cite{kamnitsas17}, however this strategy does little for the class imbalance when large patches are being considered, as individual patches tend to still be highly class imbalanced at a voxel level. Another option, is to sample patches centered at WMH locations only, which in fact reduces the healthy tissue class to $\sim$90\%. However, both strategies, in combination with the proposed architecture that has a field of view comparable to sample size, would lead to a location bias, where WMHs are always expected in the center of the patch. Instead, we propose that after defining a random subset of WMH voxels from which to extract training patches, a random shift $\Delta_{x,y}$ of up to half the patch size be applied in the axial plane before patch extraction to augment the dataset. Figure \ref{fig:patch_sampling} details this procedure. It is important to point out that the location sensitivity mentioned here, is generally not an issue with dense training in natural images, where different classes can either appear anywhere in a scene (e.g. a face might be located anywhere), or class location gives a meaningful description (e.g. sky tends to be in the upper part of a scene). This problem only occurs when sampling patches from training images in a systematic way, such as proposed here.

\section{Data}\label{sec:data}
The proposed methodology was evaluated using a subset of 167 images from 250 consecutive patients who presented themselves to a hospital stroke service with their first clinically evident non-disabling lacunar or mild cortical ischemic stroke \cite{valdeshernandez15a}. Diabetes, hypertension, and other vascular risk factors were not criteria for exclusion. However, patients with unstable hypertension or diabetes, other neurological disorders, major medical conditions including renal failure, contraindications to MRI, unable to give consent, those who had hemorrhagic stroke, or those whose symptoms were resolved within 24 hours (i.e., transient ischemic attack) were excluded. The subset of 167 subjects considered in this work consisted of those for which all WMH and stroke lesions were delineated (see Section \ref{ssec:gt}) as different annotation classes, i.e. those that contained strokes but were not labeled as such were excluded. In this work, stroke lesions included both old and recent lesions as defined in \cite{valdeshernandez15a}, which in turn are either of cortical or sub-cortical nature.

A subset of 126 from the 167 subjects used, contained additional complete clinical and demographic data. Information included risk factors and clinical assessments such as: age, sex, reported diabetes, reported hypertension, reported hyperlipidaemia, reported smoking, mini mental state examination (MMSE), systolic blood pressure (SBP), diastolic blood pressure (DBP), total cholesterol, peri-ventricular Fazekas score (PV-Fazekas), deep white matter Fazekas score (D-Fazekas), deep atrophy volume (deepAtrophyVol), basal ganglia enlarged peri-vascular spaces (BGPVS) score, centrum semiovale enlarged peri-vascular spaces (CSPVS) score, old stroke lesion (oldLes) present, and total number of micro-bleeds (micrBld).

\subsection{MRI acquisition}
All image data was acquired at the Brain Research Imaging Centre of Edinburgh (http://www.bric.ed.ac.uk) on a GE Signa Horizon HDx 1.5T clinical scanner (General Electric, Milwaukee, WI), equipped with a self-shielding gradient set and manufacturer-supplied eight-channel phased-array head coil. Details of the protocols used for acquiring the data are given in Table \ref{tab:MR_protocol}, and their rationale is explained in \cite{valdeshernandez15a}. Although several imaging sequences were acquired, only T1 and FLAIR MR images were used for this study. Of the 167 subjects considered in this work 35 were acquired under protocol 1, 83 under protocol 2 and 49 under protocol 3.

\begin{table}[ht]
\begin{center}
{\footnotesize
\begin{tabular}{|c|ccc|}
\hline \hline
\textbf{Protocols}						&\textbf{Protocol 1}      		&\textbf{Protocol 2}  	&\textbf{Protocol3} 	\\ \hline \hline
TR/TE/TI (ms) T1						&9/440			  				&\multicolumn{2}{c|}{9.7/3.984/500}     	\\ \hline\hline
TR/TE/TI (ms) FLAIR						&9002/147/2200		  			&\multicolumn{2}{c|}{9000/140/2200}     	\\ \hline
\multirow{2}{*}{Pixel bandwidth (KHz)} 	&125 (T1)		  				&\multicolumn{2}{c|}{15.63 (T1)}    	\\ 
										&122.07(FLAIR)		  			&\multicolumn{2}{c|}{15.63 (FLAIR)} 	\\\hline
\multirow{2}{*}{Matrix} 				&\multirow{2}{*}{256x192} 		&256x216 (T1)  			&192x192 (T1) \\
										&			  					&384x224 (FLAIR) 		&256*256(FLAIR) \\\hline
\multirow{2}{*}{No. slices}				&\multirow{2}{*}{20}	  		&256 (T1)				&160 (T1) \\
										&			  					&28 (FLAIR)				&40 (FLAIR)\\\hline
\multirow{2}{*}{Slice thickness (mm)}   &\multirow{2}{*}{5}	  			&1.02 (T1)				&1.3 (T1)\\
				        				&			  					&5 (FLAIR)				&4 (FLAIR)\\\hline
Inter-slice gap (mm) 					&1.5			  				&1						&0\\
\multirow{2}{*}{Voxel size (mm$^3$)}	&\multirow{2}{*}{0.94x0.94x6.5}	&1.02x0.9x1.02 (T1)		&1.3x1.3x1(T1)\\
										&			  					&0.47x0.47x6 (FLAIR)	&1x1x4 (FLAIR)\\
\hline
\end{tabular}}
\caption{MR imaging sequence details for the three acquisition protocols used.}
  \label{tab:MR_protocol}
\end{center}
\end{table}

\subsubsection{Image pre-processing and gold standard annotations} \label{ssec:gt}
All image sequences (from each patient) were co-registered using FSL-FLIRT \cite{jenkinson02} and mapped to the patient’s FLAIR space.  
WMH from MR images that were acquired under protocol 2 (Table \ref{tab:MR_protocol}) were delineated using Multispectral Coloring Modulation and Variance Identification (MCMxxxVI). Described in \cite{valdeshernandez15a,valdeshernandez15b}, MCMxxxVI is based on the principle of modulating, or mapping, in red/green/blue color space two or more different MRI sequences that display the tissues/lesions of the brain in different intensity levels, before employing Minimum Variance Quantization (MVQ) as the clustering technique to segment different tissue types. Here, MCMxxxVI considers WMH those hyperintense signals that simultaneously appear in all T2-weighted based sequences. 
WMH from MR images acquired under the protocols 1 and 3 were delineated via a human corrected histogram-based threshold of the FLAIR sequence. Stroke lesions (old and recent) were separately extracted semi-automatically by thresholding and interactive region-growing, while guided by radiological knowledge, on FLAIR (if ischemic) or T2*W (if hemorrhagic) \cite{valdeshernandez15a,valdeshernandez15b}.
Their identification procedure is described in the Table 2 of \cite{valdeshernandez15a} and single stroke class was created by combining recent and old. All images were re-sliced as to have $1mm$ in both dimensions of axial slices, with the remaining dimension (slice thickness) left unchanged. White matter probability maps were obtained from T1 image segmentation using \cite{ledig15} and cerebro-spinal track probability maps were obtained by co-registering a tract probability map \cite{hua08} to the FLAIR image space. Additionally, in order to have consistent intensity voxel values for model training all MR images were normalized as to have zero mean and standard deviation of one (excluding the background). Values below three standard deviations from the mean clipped in order to guarantee consistent background values across all images.

\section{Experiments and results} \label{sec:results}
Data used as input for training the proposed CNN (uResNet) was sampled from whole brain MR volumes as explained in Section \ref{sec:sampling}. Image patches and the corresponding labels of $64 \times 64$ voxels were extracted from the volumes at a random subset of 20\% of all possible locations that were labeled as WMH and 80\% of locations labeled as stroke. All 167 images included in this study contained WMH lesions, of these, 59 also contained stroke lesions. Data was split into two separate sets used for two fold cross-validation, where each fold contained half of the images with WMH only and half with both WMH and stroke, as to represent data distribution in both folds. 
During each fold of the cross validation experiments one fold is used for training (network parameter learning) and setting all other parameters, while the second (unseen) fold is reserved for testing. That is, optimization of the loss function, input channel selection and stopping criteria are carried out on the training set.
\ref{sec:appA} for a comparison of the proposed uResNet with a version that used residual blocks with two convolutions (uResNet2) and to observe the added value of the center shifting during training patch sampling (uResNet\_NoC).
Experiments were  carried out using the Theano \citep{theano} and Lasagne \cite{lasagne} frameworks with Adam \cite{kingma14} optimization (default Lasagne parameters), mini-batch size of 128, learning rate of 0.0005 (with 0.1 decay after 25 epochs) and random weight initialization (default Lasagane settings).

The evaluation criteria used to compare all methods can be split in two, mainly, a comparison to other well established and state-of-the-art methods and a clinical analysis. The comparison to other methods consisted of an evaluation of the labeling overlap of the segmented images using the Dice coefficient, and an analysis of the differences between the automatically calculated volumes and the expert in terms of intra-cranial volume (ICV). Comparison results calculated using the Dice coefficient and volume analyses are reported on a per class basis. Clinical evaluations consisted of a correlation analysis with some clinically relevant variables (mainly Fazekas and MMSE scores), and a general linear model (GLM) analysis of association with known risk factors.

\subsection{Model training}
\begin{figure}[t]
\centering
\includegraphics[width=12cm]{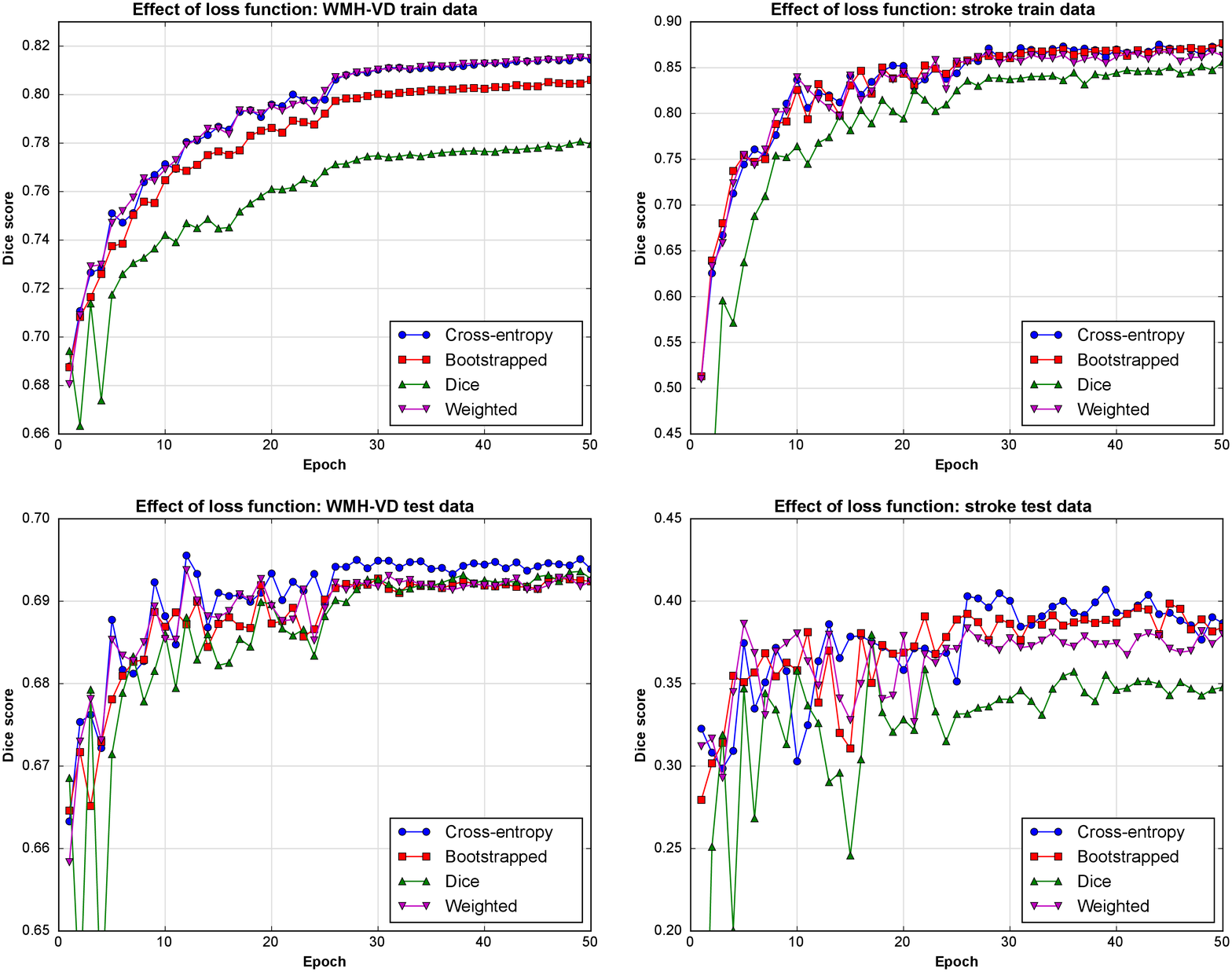} 
\caption{Effect of different loss functions on uResNet trained using FLAIR images.}
\label{fig:dice_loss}
\end{figure}

An important factor during CNN training is the definition of the loss function that will guide the learning process (Section \ref{ssec:loss}). Here, we experimented with several recently proposed loss functions that were used to train the proposed WMH segmentation CNN using FLAIR images as input. In order to directly compare the effect of different loss functions, Dice score results from evaluating the CNN after different stages of training were calculated, see Figure \ref{fig:dice_loss}. Here, the horizontal axis indicates number of training epochs while the vertical axis indicates the Dice score achieved on either the train (top row) or test (bottom row) datasets. In this work, an epoch is defined as transversing the complete set of training of patches once. It must also be noted that Dice results displayed here are calculated on the whole brain MR volumes, not on the extracted patches. Figure \ref{fig:dice_loss} shows the results obtained using classical, bootstrapped and weighted cross-entropy loss functions, as well as using a pseudo Dice similarity score (see Section \ref{ssec:loss}). From the top row of Figure \ref{fig:dice_loss} (results on train data) it can be observed that weighted and classical cross-entropy perform best and that there is little difference between them. However, weighted cross-entropy has an additional class weight parameter associated that also need to be set. Hence, for the problem presented in this work and considering the experiments conducted, classical cross-entropy was considered the best choice. It is important to take notice that using the Dice coefficient as both loss function and evaluation metric provides surprisingly poor results during training (top row Figure \ref{fig:dice_loss}). Here, we theorize that, for this particular problem, the solution space over which we optimize might be more complex for the Dice metric than the other, and hence finding an global optimal solution might prove more cumbersome.

\begin{figure}[t]
\centering
\includegraphics[width=12cm]{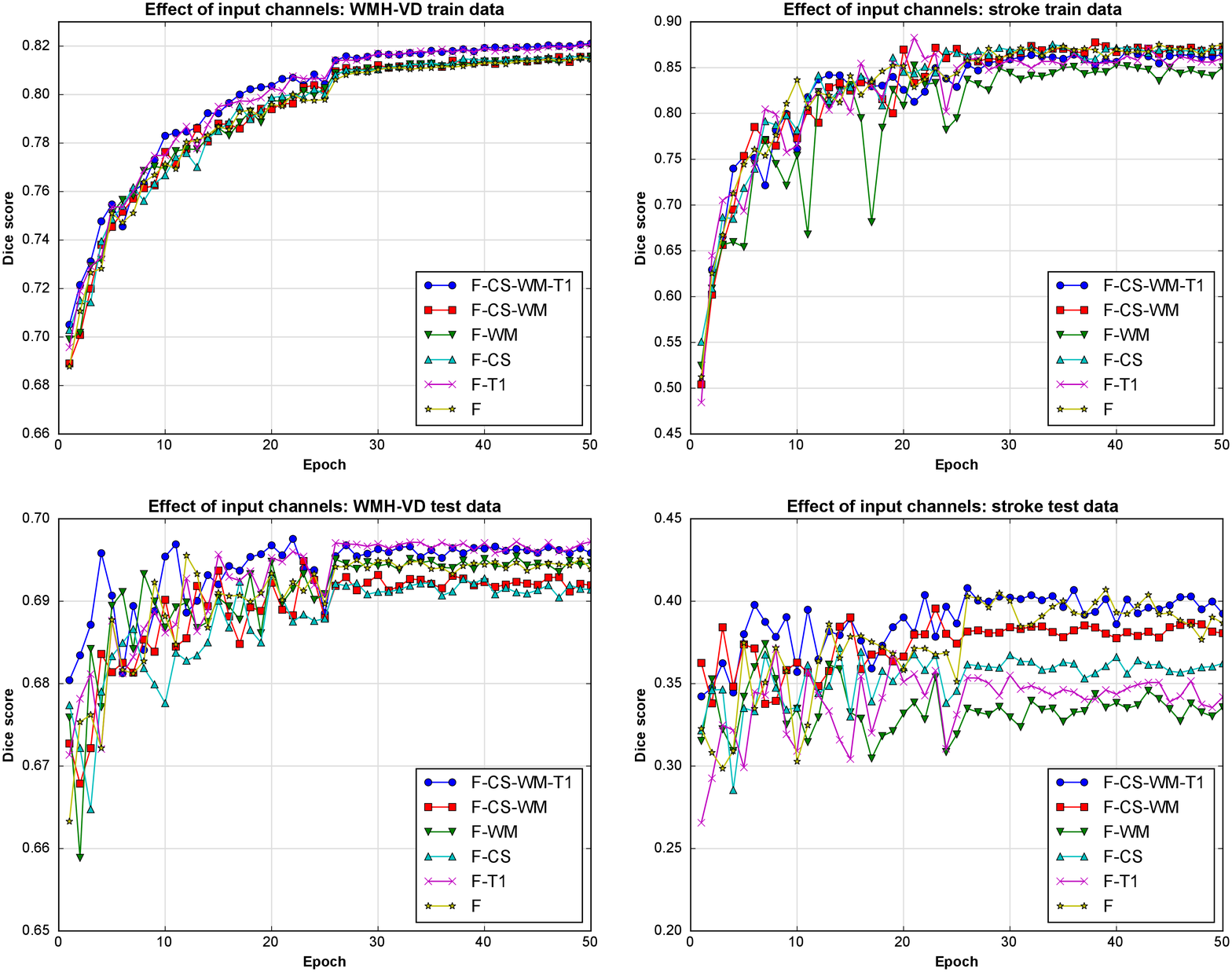} 
\caption{Different input channel exploration. F: FLAIR image, CS: cerebro-spinal track atlas, WM: white matter probability map, T1: T1 weighted image.}
\label{fig:dice_channels}
\end{figure}

As mentioned before, WMHs are best observed in FLAIR MR images, however it has been suggested that complementary information might be found on T1 MR images. In this work, the contribution from additional multi-modal information to the proposed segmentation framework was explored. Additional input channels to the proposed CNN include T1 MR images, white matter probability maps and a cerebro-spinal tract atlas. Segmentation accuracy is again evaluated using the Dice score. From Figure \ref{fig:dice_channels}, it can be seen that training converges after about 30 epochs, that is, traversing the whole set of extracted training patches 30 times. Therefore, test Dice scores and automatic volumes further presented here are obtained evaluating the model at 30 epochs.

Given the different input channels, training data and testing results that take into account both segmentation classes (shown in Figure \ref{fig:dice_channels}) indicate that there is very little difference between using four input channels (FLAIR, T1, WM and CS) compared to just using FLAIR images. Hence, all subsequent experiments made use of only FLAIR images as input channels. This is additionally justified by the fact that some of the comparison methods only use FLAIR images. Furthermore the acquisition of additional modalities (T1) or probability map generation (WM and CS) can be costly/time consuming and render the methodology less clinically viable. In Figures \ref{fig:dice_loss} and \ref{fig:dice_channels} it can also be observed that training and testing Dice scores for stroke segmentations are much more oscillatory than those from WMH segmentation. This behavior can be explained by the fact that there is simply a lot less data of the stroke class, in fact there are $\sim$14 times more WMH voxels. Therefore, stroke results are more sensitive to variations in the network's parameters as each epoch provides more stochastic gradients associated to this class. Furthermore, the stroke higher training accuracy combined with the lower test accuracy can be attributed to this class imbalance as they potentially point to an over-fitting problem.

\subsection{Comparison to state-of-the-art}
In the experiments presented in this section the proposed uResNet segmentation CNN was compared to other well established and state-of-the-art algorithms. From the lesion segmentation toolbox (LST) version 2.0.15 (http://www.statistical-modelling.de/lst.html) the LPA and LGA frameworks were used. LPA was used using only FLAIR images as input while LGA required both FLAIR and T1 images. DeepMedic, a recently published CNN library for segmentation of medical images, was also used in the comparisons presented here with its default settings. Parameters for both LPA and LGA frameworks were set according to a two fold cross-validation using the same data splits as described before for uResNet. LPA has only one parameter, a threshold $\tau$ used to binarize the lesion probability maps generated, and the optimal value $\tau$ after cross-validation was set to 0.16. The authors recommend setting this value to 0.5, however this produced poor results and hence were excluded from further analysis. LGA has two parameters, $\kappa$ that was set to 0.12 after cross-validation and a threshold that was set to 0.5. 
DeepMedic was also validated using the same two fold cross-validation strategy (with FLAIR images as input), where the network is trained in one fold and tested in the other, however, no other meta-parameter (e.g. network's filter sizes, number of feature maps or learning rate) tuning was done. 
DeepMedic was trained using images re-sliced to isotropic 1$mm^3$ voxel size, and patch sampling was internally handled by DeepMedic. The default sampling option was used, which randomly samples 50\% of patches from healthy tissue and 50\% from pathological tissue (without considering different class weightings).

\begin{figure}[t]
\centering
\includegraphics[width=12cm]{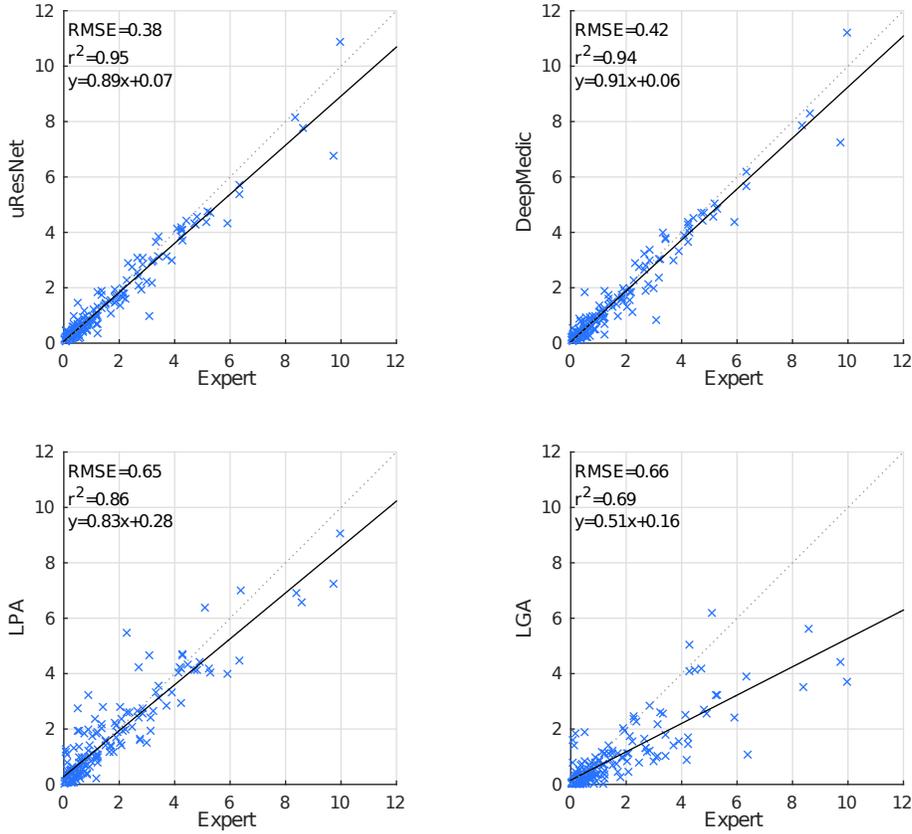} 
\caption{Automated versus expertly generated WMH volumes (as ICV \%). The solid line indicates the linear trend $f(x)$ of the comparison, while the dotted line indicates the ideal trend $f(x)=1.0x+0.0$.}
\label{fig:aut_man_WMH}
\end{figure}

Dice overlap scores between automatically generated and expertly annotated WMH and stroke lesions are shown in Table \ref{tab:mean_WMH_dice}. Here, it can be observed that the proposed uResNet outperforms the compared methods, with all comparisons between the Dice scores obtained with the proposed and every competing being found to be statistically significant $p<0.01$ according to Wilcoxon's signed rank test.
Statistical significance gives a measure of the likelihood that the difference between two groups could be attributed to change, while effect size (or the ``strength of association'') quantifies the relative magnitude of the difference between those two groups. Cohen \cite{cohen88p40} describes effect size values of 0.2, 0.5 and 0.8 as small, medium and large, respectively.
Effect sizes related to the statistical significance tests were calculated with $z/\sqrt[]{n_1 + n_2}$ as suggested by Pallant \cite{pallant10}, and were 0.45, 0.32 and 0.61 for the comparison of uResNet Dice scores against those from DeepMedic, LPA and LGA, respectively. Figure \ref{fig:aut_man_WMH} shows a correlation analysis between the expertly annotated WMH volumes and those automatically generated. To remove any potential bias associated with head size and thus allow a better comparison, volumes were converted to ICV \%. Ideally, automatic algorithms should produce values as similar as possible to the expert, and hence, should lie close to the dotted lines in Figure \ref{fig:aut_man_WMH}. The solid lines indicate the general linear trend of the expert vs. automatic comparison and the coefficient of determination $R^2$ indicates to what degree automatic values explain the expert ones. From Figure \ref{fig:aut_man_WMH} bottom row we can see that both LPA and LGA perform clearly worse than the CNN approaches (uResNet and DeepMedic). It is also evident that LPA outperforms LGA, where each has a $R^2$ value of 0.86 and 0.69, respectively. Differences between uResNet and DeepMedic (top row of Figure \ref{fig:aut_man_WMH}) are less evident. However, on close inspection of the $R^2$ metric in Table \ref{tab:mean_WMH_dice} of uResNet and DeepMedic we can see that uResNet results are slightly better correlated to those generated by the expert. On the other hand, DeepMedic has a slope of 0.91 (offset 0.06) while uResNet has a slope of 0.89 (offset 0.07), suggesting a slightly better agreement.

\begin{figure}[t]
\centering
	\includegraphics[width=12cm]{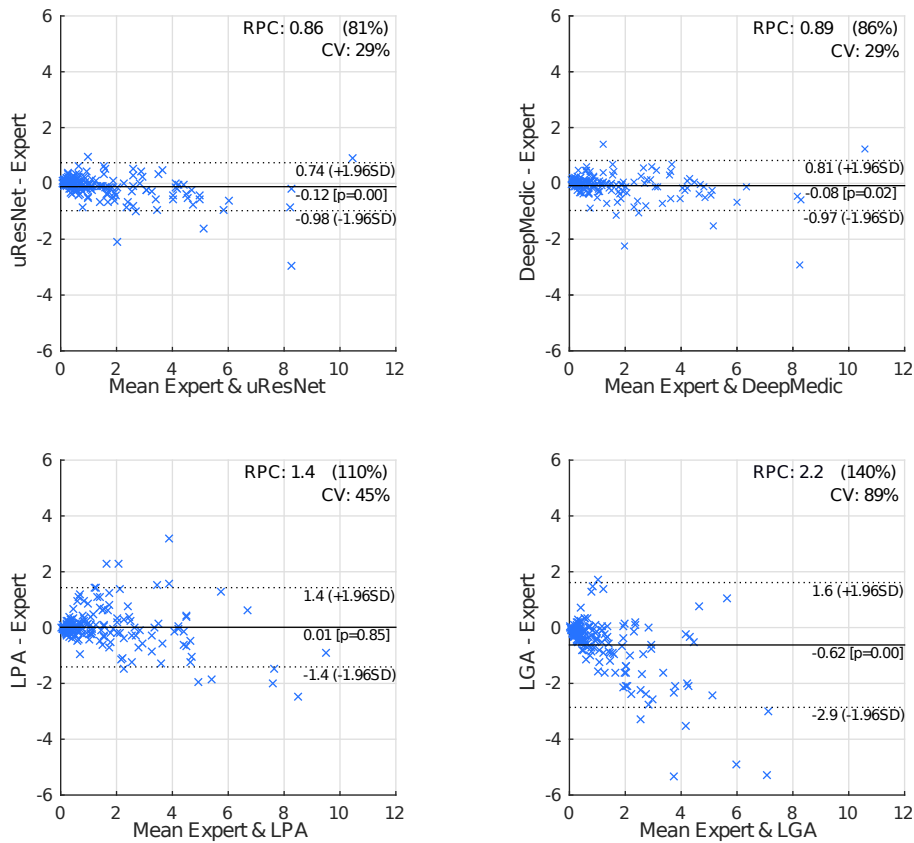} 
\caption{Bland-Altman plots comparing expert annotations with all other methods in WMH segmentation.}
\label{fig:BA_all_WMH}
\end{figure}

Figure \ref{fig:BA_all_WMH} shows Bland-Altman plots that further compare expert and automatic WMH volumes. In these plots, the horizontal axis gives the average between expert and automatic volumes for each subject, while the vertical axis shows the difference between these volumes. 
The reproducibility coefficient (RPC), as calculated here, gives a measure of the variability (or spread) of the differences between automatic and manual volumes and is calculated as 1.96 times the standard deviation $\sigma$ of those differences ($1.96*\sigma$). In the experiments presented here, smaller values indicate better agreement between automatic and manual volumes. The coefficient of variation (CV) is given by $100*\sigma/\bar{X}$, where $\bar{X}$ refers to the mean volume from both measurements.
Dotted lines in the plots of Figure \ref{fig:BA_all_WMH} give the range of the RPC. Bland-Altman plots also provide insight into possible biases of compared methods. LGA displays a statistically significant ($p=0.85$ to reject zero mean hypothesis) tendency to under-estimate volumes (central solid line). However, all methods tend to under-estimate larger volumes and over-estimate small ones, with the effect more pronounced in LGA.

\begin{figure}[t]
\centering
\includegraphics[width=12cm]{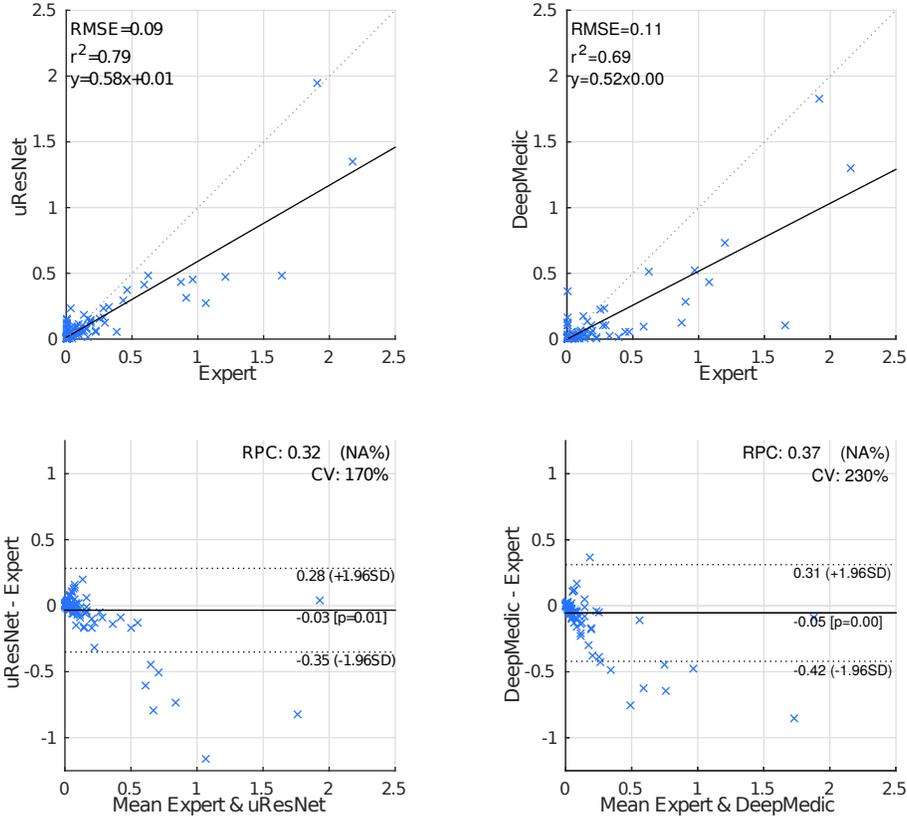} 
\caption{Automated versus expertly generated stroke volumes. LPA and LGA are unable to distinguish between WMH and stroke, hence cannot generate these results. The solid line indicates the linear trend $f(x)$ of the comparison, while the dotted indicates the ideal trend $f(x)=1.0x+0.0$.}
\label{fig:BA_corr_all_stroke}
\end{figure}

One of the main objectives of the work presented here is to also differentiate between WMH and stroke lesions. Neither LPA or LGA are capable of making such a distinction, and therefore are not suitable algorithms for this problem. Figure \ref{fig:BA_corr_all_stroke} (top-row) shows the correlation analysis between automatic (uResNet and DeepMedic) and expert stroke volumes (normalized as ICV). It is evident that uResNet outperforms DeepMedic in terms of RMSE, $R^2$ and linear fit slope. Further to this analysis, Figure \ref{fig:BA_corr_all_stroke} (bottom-row) shows Bland-Altman plots that further confirm these findings, where uResNet obtains a smaller RPC and CV than DeepMedic, with neither method on average displaying a statistically significant tendency to over- or under-estimate volumes (see central solid line on plots). However, it is worth noting that both methods have a tendency to over-estimate small volumes and under-estimate larger ones. A summary of Figures \ref{fig:aut_man_WMH} and \ref{fig:BA_corr_all_stroke} is also presented in Table \ref{tab:mean_WMH_dice}, where a difference between both algorithms in terms of Dice scores can be observed. Statistical significance between the comparison of uResNet and DeepMedic Dice scores was found to be $p<0.05$ according to Wilcoxon's signed rank, with an effect size related to this statistical significance (as suggested by \cite{pallant10}) of 0.12.
The gap between uResNet and DeepMedic can be considerably closed if additional inputs are provided to DeepMedic (see \ref{sec:app}), however this requires an additional MR image acquisition (and co-registration of such image), tissue segmentation and/or co-registration of a cerebro-spinal track atlas. Furthermore, 
in \ref{sec:appC} results of
DeepMedic experiments that aim to approximate the sampling scheme used by uResNet are discussed.

\begin{sidewaysfigure}
    \centering
    \includegraphics[width=21cm]{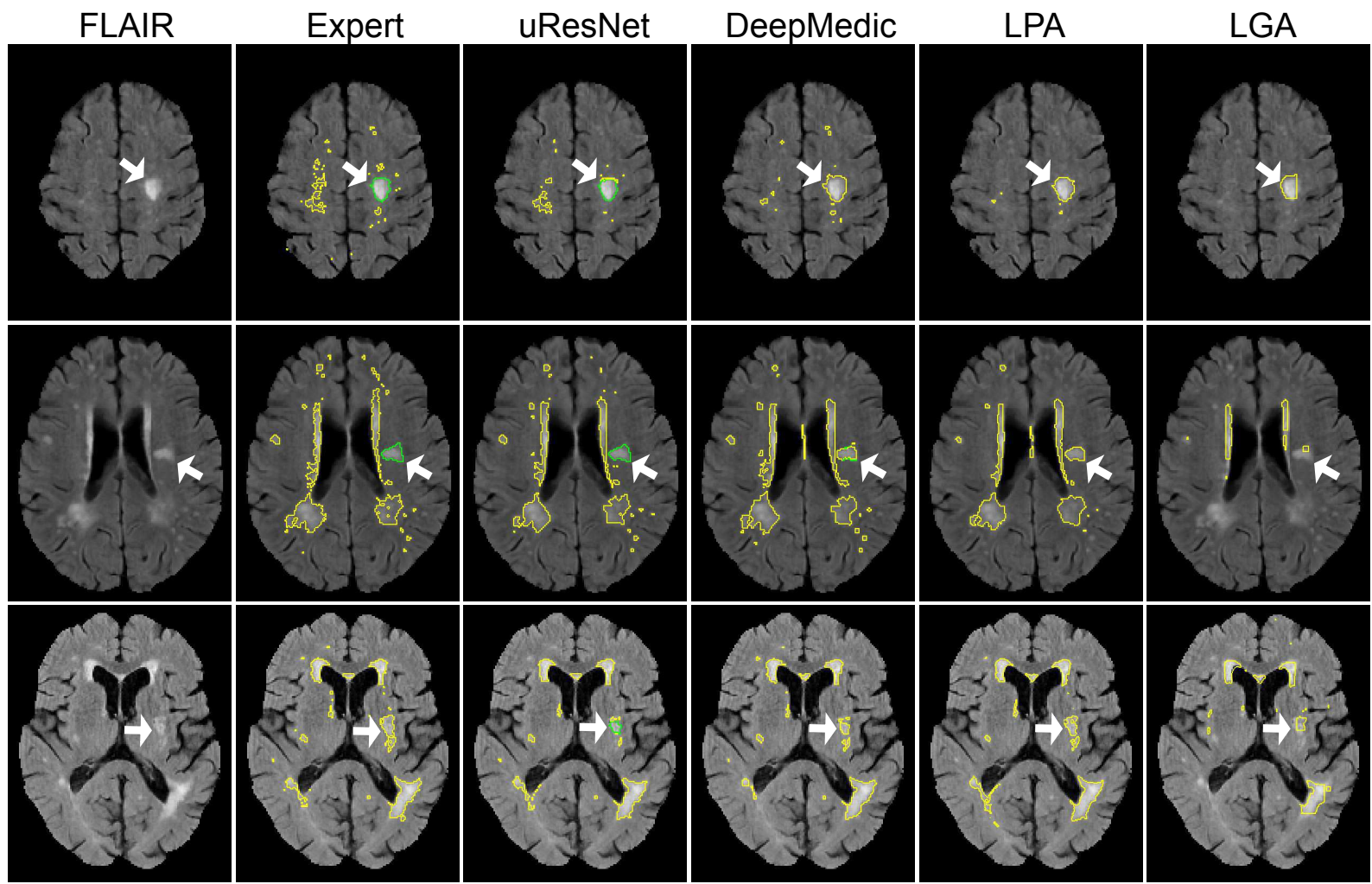} 
    \caption{Visual comparisons of all competing methods. Yellow lines delineate WMH, green lines stroke and white arrows point to interesting result areas. Best seen in color.}
    \label{fig:visual_results}
\end{sidewaysfigure}

Figure \ref{fig:visual_results} shows the segmentation results from three example subjects that illustrate the differences between the methods. Here, it can be observed that uResNet generally does a better job at differentiating between WMH and stroke lesions when compared to DeepMedic (top and middle row). In the bottom row of Figure \ref{fig:visual_results} and example is illustrated when uResNet wrongly segments some WMH as stroke. Additionally, in the top row,
all methods are shown to clearly under-segment the image when compared to the expert is shown. However, inspecting the FLAIR image of this subject (top row, leftmost column) it can be seen that the under-segmented regions would be challenging even for another expert annotator.

\subsection{Clinical evaluation}
Experiments thus far indicate a better agreement between volumes generated by uResNet and expert annotations, however, the question of the clinical validity of such results remains open. In this regard, Table \ref{tab:mean_WMH_dice} gives correlation coefficient (CC) results between the volumes and some clinical variables (Fazekas scores and MMSE). Fazekas scores were split into deep white matter (D-Fazekas) and peri-ventricular (PV-Fazekas), with values ranging from 0-3. An additional combined Fazekas score, created by adding both D-Fazekas and PV-Fazekas, is also presented. From Table \ref{tab:mean_WMH_dice} we can observe that in terms of correlation to Fazekas score the proposed uResNet outperforms the other competing methods, additionally noting that CC results for PV-Fazekas and Fazekas are even higher than those obtained from the expert annotations. However, in terms of CC with MMSE it was LPA that performed best.

\begin{table}
\begin{center}
{\footnotesize
\renewcommand{\tabcolsep}{5pt}
\begin{tabular}{ |l|cccc|c| }
\hline \hline
					&\textbf{uResNet}	&\textbf{DeepMedic}	&\textbf{LPA}	&\textbf{LGA}	&\textbf{Expert}\\\hline \hline
WMH Dice (std)		&\textbf{69.5(16.1)}&66.6(16.7)			&64.7(19.0)		&41.0(22.9)		&-				\\ 
Stroke Dice (std)	&\textbf{40.0(25.2)}&31.3(29.2)			&-				&-				&-				\\ 
WMH $R^2$			&\textbf{0.951}		&0.943				&0.855			&0.687			&-				\\ 
Stroke $R^2$		&\textbf{0.791}		&0.688				&-				&-				&-				\\ 
WMH Trend			&0.89x+0.07			&\textbf{0.91}x-\textbf{0.06}		&0.83x+0.28		&0.51x+0.16		&-				\\ 
Stroke Trend		&\textbf{0.58x}+0.01&0.52x-\textbf{0.00}&-				&-				&-				\\ 
CC D-Fazekas 		&\textbf{0.770}		&0.769				&0.746			&0.630			&0.774			\\ 
CC PV-Fazekas		&0.778				&\textbf{0.780}		&0.777			&0.718			&0.765			\\ 
CC Fazekas			&\textbf{0.824}		&\textbf{0.824}		&0.811			&0.734			&0.819			\\ 
CC MMSE				&0.364				&0.369				&\textbf{0.443}	&0.389			&0.372			\\ 
\hline
\end{tabular}}
\caption{Mean Dice scores of WMH and stroke (standard deviation in parenthesis), correlation analysis between expert and automatic volumes ($R^2$ and trend), and correlation with clinical variables.}
  \label{tab:mean_WMH_dice}
\end{center}
\end{table}

Using the clinical scores as well as known risk factors available, an analysis of association between WMH volumes and risk factors was carried out. In order to explore such associations a GLM between the results of every algorithm (as well as the expert) and the risk factors was generated. In these models, the risk factors and clinical scores were treated as dependent variables, while the volumes acted as the independent variable. After careful consideration, age, sex, reported diabetes, reported hypertension, reported hyperlipidaemia, reported smoking, total cholesterol, deep atrophy volume and BGPVS score were used in the GLM analysis. Table \ref{tab:associations} provides p-values that indicate if a particular risk factor associated with the generated WMH volumes, where the GLMs were corrected for gender differences. Results indicate that only BGPVS is found to be associated with the expertly generated volumes, however deep atrophy volume was also found to be associated with all other methods. Additionally, LPA volumes were also found to be associated with age and diabetes.

\begin{table}
\begin{center}
{\footnotesize
\renewcommand{\tabcolsep}{5pt}
\begin{tabular}{ |l|cccc|c| }
\hline \hline
				&uResNet			&DeepMedic			&LPA				&LGA				&Expert		\\ \hline \hline
age				&0.491				&0.533				&$<$\textbf{0.001}	&0.723				&0.313		\\ 
diabetes		&0.082				&0.072				&\textbf{0.003}		&0.070				&0.066		\\ 
hyperlipidaemia	&0.645				&0.547				&0.551				&0.687				&0.728		\\ 
hypertension	&0.820				&0.781				&0.504				&0.358				&0.562		\\ 
smoking			&0.497				&0.560				&0.216				&0.719				&0.767		\\ 
totalChl		&0.235				&0.281				&0.161				&0.328				&0.371		\\ 
BGPVS			&$<$\textbf{0.001}	&$<$\textbf{0.001}	&$<$\textbf{0.001}	&$<$\textbf{0.001}	&$<$\textbf{0.001}		\\ 
deepAtrophyVol	&\textbf{0.015}		&\textbf{0.019}		&$<$\textbf{0.001}	&$<$\textbf{0.001}	&0.117		\\

\hline
\end{tabular}}
\caption{P-values of linear regression associations between volumes calculated with different methods and risk factors. Bold numbers indicate
statistical significance above 0.05.}
  \label{tab:associations}
\end{center}
\end{table}

\begin{figure}
\centering
\includegraphics[width=12cm]{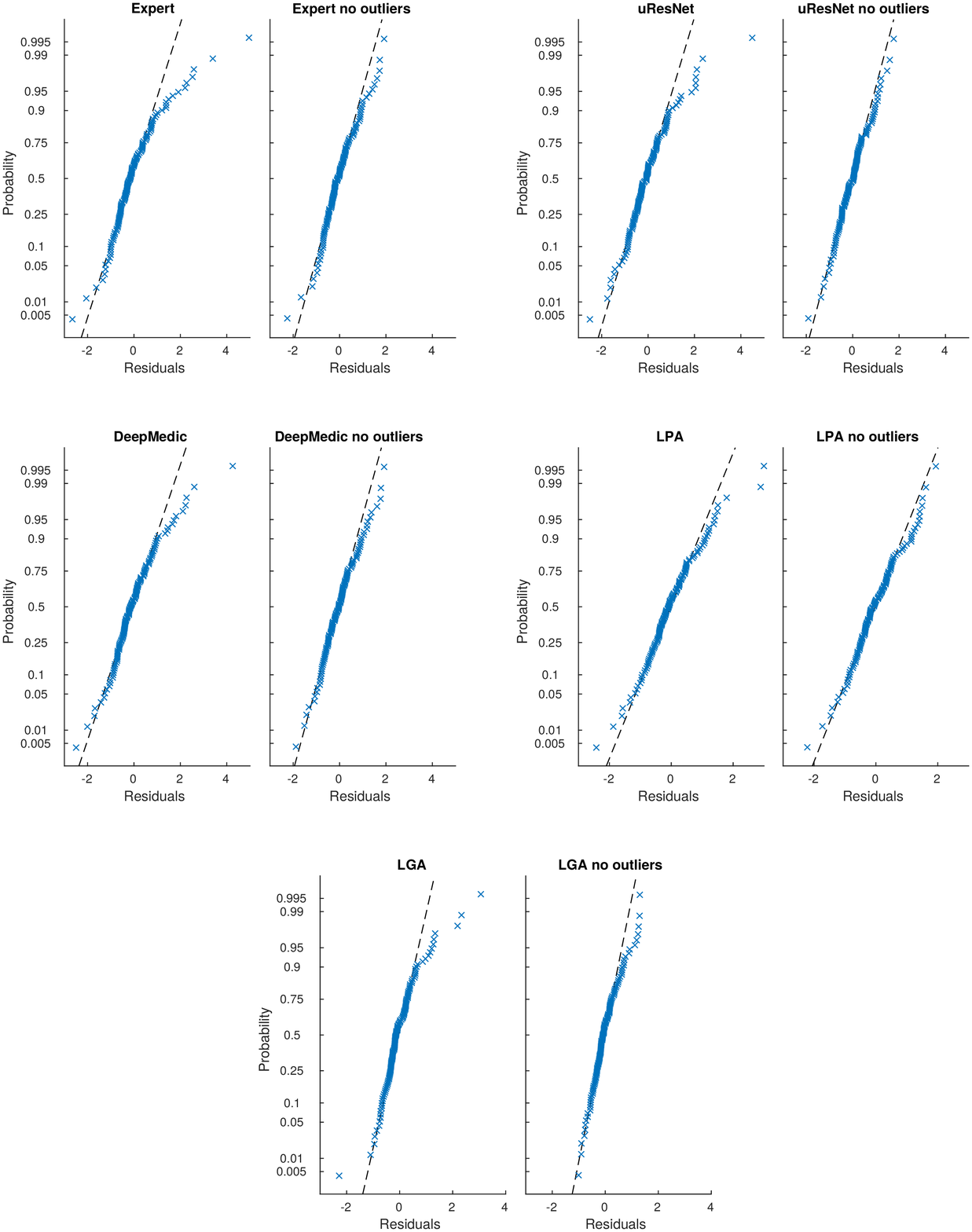} 
\caption{General linear model normal probability plots of residuals for all methods, with and without outliers.}
\label{fig:nppr}
\end{figure}

In GLM analysis, values that are not well described by the model (outliers) can have a significant impact in subsequent analyses. Outliers in GLM can be identified by examining the probability distribution of the residuals. In order to eliminate any potential bias introduced by outliers, an analysis with outliers removed was performed. Results of this outlier-free association analysis are presented in Table \ref{tab:associations_no_outliers}. Figure \ref{fig:nppr} shows the normal probability plot of residuals for all methods before and after of outlier removal. From Table \ref{tab:associations_no_outliers} we can observe that once outliers were removed, expert volumes were found to be associated with deep atrophy volume, BGPVS and diabetes. The same associations were found for uResNet, DeepMedic and LPA, with the addition that LPA was again also associated with age. LGA was found to only be associated with BGPVS and deep atrophy volume.

\begin{table}
\begin{center}
{\footnotesize
\renewcommand{\tabcolsep}{5pt}
\begin{tabular}{ |l|cccc|c| }
\hline \hline
					&uResNet			&DeepMedic			&LPA				&LGA				&expert		\\ \hline \hline
age					&0.905				&0.993				&$<$\textbf{0.001}	&0.685				&0.407		\\ 
diabetes			&\textbf{0.012}		&\textbf{0.019}		&$<$\textbf{0.001}	&0.177				&\textbf{0.003}		\\ 
hyperlipidaemia		&0.346				&0.425				&0.464				&0.550				&0.186		\\ 
hypertension		&0.639				&0.502				&0.190				&0.128				&0.350		\\ 
smoking				&0.069				&0.084				&0.107				&0.673				&0.343		\\ 
totalChl			&0.294				&0.212				&0.222				&0.043				&0.868		\\ 
BGPVS				&$<$\textbf{0.001}	&$<$\textbf{0.001}	&$<$\textbf{0.001}	&$<$\textbf{0.001}	&$<$\textbf{0.001}		\\ 
deepAtrophyVol		&\textbf{0.005}		&\textbf{0.008}		&$<$\textbf{0.001}	&$<$\textbf{0.001}	&\textbf{0.020}		\\ 

\hline
\end{tabular}}
\caption{P-values of linear regression associations between volumes calculated with different methods and risk factors after residual outliers were removed. Bold numbers indicate statistical significance above 0.05.}
  \label{tab:associations_no_outliers}
\end{center}
\end{table}

Fazekas scores are highly co-linear with WMH volume (the dependent variable) and therefore were excluded from all previous GLM analysis. Nonetheless, a GLM that included Fazekas scores was also composed as a sanity check that the correct associations would be found. A single Fazekas score was generated by adding the D-Fazekas and PV-Fazekas scores (0-6 score scale). All models found a very strong association ($p\ll0.001$) between Fazekas and WMH volumes. The effect size for the association of Fazekas score with the expertly generated WMH volumes indicates that a change of one in the Fazekas scale, translates to a change of 0.75 ICV \% increase of WMHs (1-0.75). DeepMedic obtained the closest effect size of the association between Fazekas scores and WMH volumes to that of the expert, with a prediction that an increase of one Fazekas point produces 0.70 ICV \% increase of WMH (1-0.70). uResNet closely followed with 1-0.69 predictions. LPA and LGA results produced effect sizes of 1-0.6 and 1-0.35, respectively. Of the expert stroke lesion volumes, 
systolic blood pressure was the only risk factor to be found associated ($p<0.05$), which incidentally was also associated with the automatically (uResNet and DeepMedic) generated volumes. uResNet values were additionally found to be associated with hypertension.
However, it is important to note the small size and heterogeneous nature of the population used in this analysis, which might not prove sufficient to uncover some associations. Due to the small sample analyzed no outlier removal analysis was performed for stroke associations.

\section{Discussion}\label{sec:discussion}
In this work we have proposed a CNN framework, uResNet, for the segmentation of WMHs that is capable of distinguishing between WMHs arising from different pathologies, mainly WMHs of presumed VD origin and those from stroke lesions. Comparison results indicate that the proposed uResNet architecture outperforms other well established and state-of-the-art algorithms.

The architecture used in uResNet follows closely the architecture of U-Net \cite{ronneberger15a}. The main difference being the use of residual elements and a generally lower complexity through the use of summation instead of concatenation in skip connections. Preliminary experiments with both summation and concatenation of features maps found no difference in performance, hence low complexity was favored. However, it is also noted that a more general solution is given by the use of concatenation, as this would allow the network to learn which is the best way of combining the feature maps during training. Of course this additional complexity comes at the expense of a higher risk of over-fitting and a higher memory consumption. As mentioned, the use of residual units provide advantages during training, mainly improved convergence rates in our experiments. Recently, He et al. \cite{he16_b} proposed a new pre-activated residual unit, which optimizes the architecture of each unit making training easier and improving generalization. Future work will involve updating the architecture to include such residual elements and evaluating their merits in the context of WMH segmentation.

Large class imbalance in medical image segmentation is generally an issue that must be considered. Loss functions that take into account the class imbalance have the drawback that they have the additional class weighting parameter to tune. An additional complication resulting from a large class imbalance is that a lot of computational effort might be spent optimizing to perform well in large and relatively easy to classify/segment sections of an image. Bootstrapped cross-entropy attempts to focus the learning process on hard to classify parts of an image by dropping out loss function contribution from voxels that have already been classified to a good degree of certainty. However, this technique also requires the setting of an additional parameter, the threshold to consider a classifications as already good, and moreover, evaluation results indicated a performance similar to classical cross-entropy. 

A very important factor of the proposed CNN framework is the training data sampling strategy described in Section \ref{sec:sampling}. CNN training for medical imaging using patches is a somewhat standard technique that helps reduce the very large class imbalance that usually affects medical image segmentation. However, careful consideration must be given in the sampling strategy adopted for a certain architecture. As mentioned, class imbalance and lesion location within samples need to be considered. The use of the proposed sampling strategy described in Section \ref{sec:sampling} had a profound effect on the proposed uResNet, with WMH and stroke Dice scores increasing from $\sim$67 to $\sim$70 and from $\sim$29 to $\sim$40, respectively, due to this alone. Another important factor is the frequency each class is sampled. In this work we sampled at 20\% of the locations labeled as WMH while at 80\% of the locations labeled as stroke, again to try to balance classes. It is important to note that the default sampling settings of DeepMedic were used as in \cite{kamnitsas17}. 
In this default sampling strategy, DeepMedic samples equally from healthy and diseased tissues (that is without, considering frequency of different diseased classes) and furthermore does not include the central voxel offset sampling strategy used here. We believe both these factors had a significant impact in the differences between these methods, specially in the stroke lesion class.
Training data was augmented by applying random flips to training patches, however we did not find this had a clear effect on results. 

An important aspect to note is that WMH segmentation is notoriously challenging: For example, Bartko \cite{bartko91} and Anbeek et al. \cite{anbeek04} consider similarity scores of 70 to be excellent, while Landis and Koch \cite{landis77} consider scores of 40, 60 and 80 to be moderate, substantial and near perfect, respectively. With this in mind, we can consider average Dice scores for WMHs generated by the proposed uResNet, as well those from DeepMedic and LPA to all be substantial, with LGA generating only moderate results. It is important to note that LGA is at heart an unsupervised method and that data was only used to tune its $\kappa$ parameter. Only uResNet and DeepMedic are capable of distinguishing between different types of lesion, and in this regard only uResNet produced an average stroke Dice score that could be considered moderate.

\section*{Acknowledgments}
The research presented here was partially funded by Innovate UK (formerly UK Technology Strategy Board) grant No. 102167 and by the 7th Framework Programme by the European Commission (http://cordis.europa.eu; EU-Grant-611005-PredictND -- From Patient Data to Clinical Diagnosis in Neurodegenerative Diseases).
Additionally, data used in preparation of this work was obtained under funding by the Row Fogo Charitable Trust (AD.ROW4.35. BRO-D.FID3668413) for MVH, and the Wellcome Trust (WT088134/Z/09/A)..

\appendix

\section{Variations of uResNet}\label{sec:appA}
In this section we present results comparing the proposed architecture and sampling scheme, with two additional version: One where the residual block takes the more traditional form of two convolutional elements (called uResNet2) and another where the proposed center shifting sampling scheme is replaced with an standard centered patch sampling scheme (called uResNet\_NoC, for not off-centered). Table \ref{tab:uResNet_comparison} summarizes these results.

Using single convolution residual blocks was noted He et al. \cite{he16} to be equivalent to a linear projection. After experimented with residuals blocks of one and two convolutions, we observed no statistical difference ($p>0.05$) between them. However, learning the residual of these linear projections might still be simpler, thus leading to an observed faster convergence. This observations need to be interpreted with care. We believe that the Dice overlap scores that our method achieves are close to expected intra-rater variability, hence the lack of observed difference in performance between one and two convolutions in residual blocks, might come down to limitations of the data itself. 

\begin{figure}[!t]
\begin{minipage}[b]{1.0\linewidth}
\centering
\subfigure[]{\includegraphics[height=5cm]{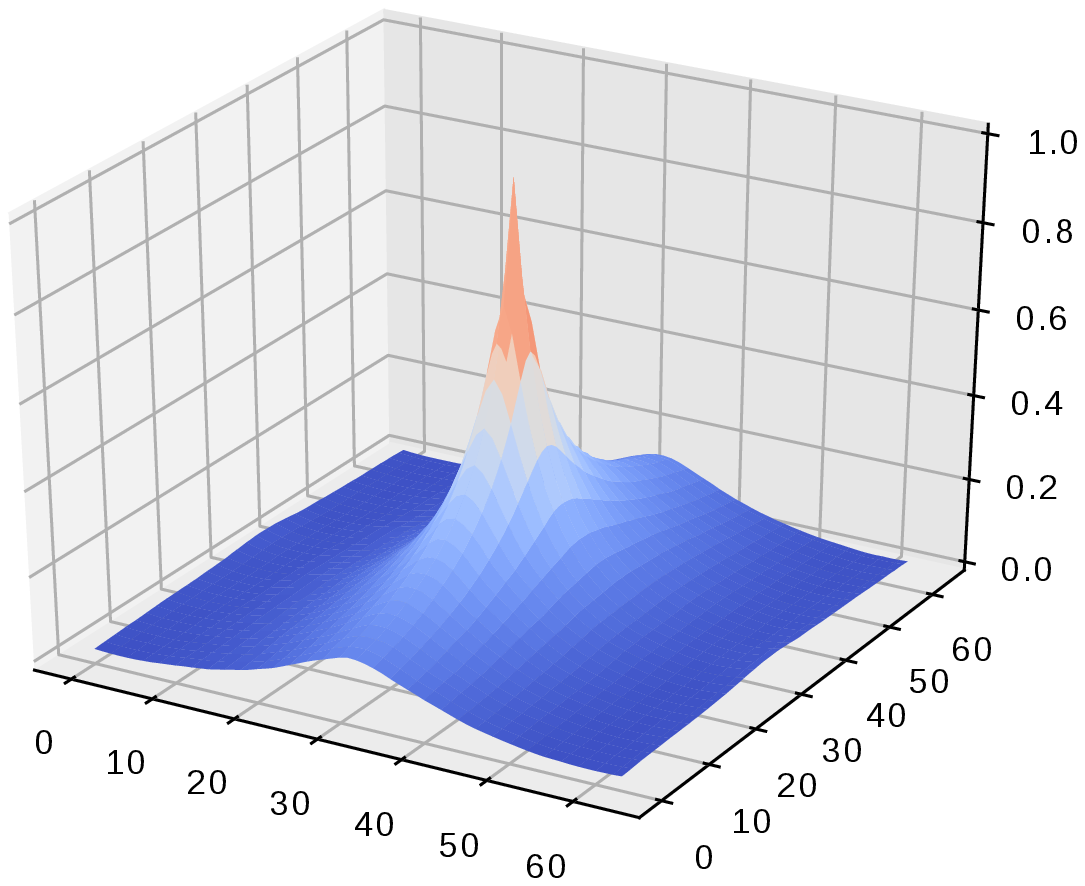}}
\subfigure[]{\includegraphics[height=5cm]{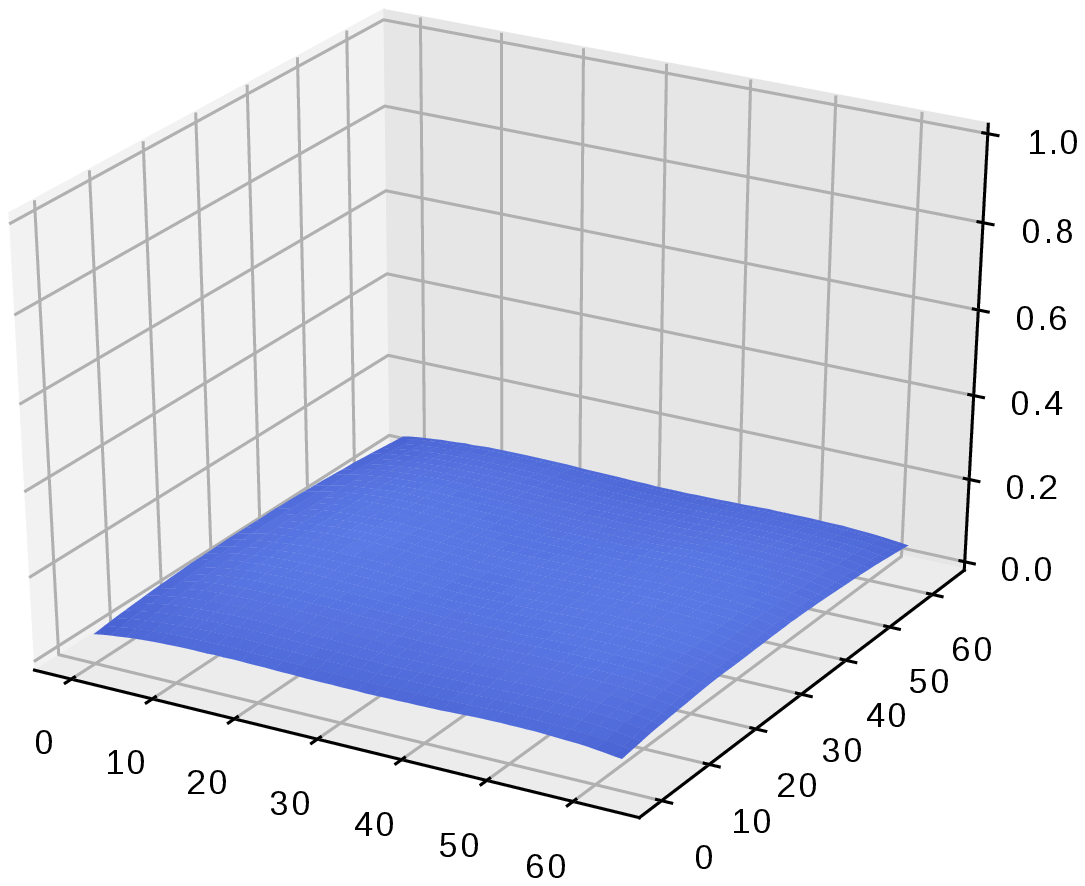}}
\end{minipage}
\caption{Lesion likelihood on training patches (a) without shifting and (b) after shifting.}
\label{fig:lesion_prob}
\end{figure}

Training with patches that always contain a diseased label in the center would bias towards labeling this region of a patch as diseased during inference. Patch center shifting alleviates this problem due to the distribution of  probability to observe a lesion across the whole field-of-view. For example, if we would estimate the probability of observing a lesion in any particular location of a training patch, there would be 100\% probability to observe a lesion at its center (Figure \ref{fig:lesion_prob} (a)), as we explicitly sampled in this manner. Allowing patches to be shifted spreads this probability to all locations and not any single location has a preferential likelihood of being a lesion (Figure \ref{fig:lesion_prob} (b)). 
In a fully convolutional neural network predictions can be made over a large area (as the network proposed here), taking into account context information from large areas of an image (the field-of-view or receptive field). However, training is driven by pixel-wise prediction errors, hence labeling occurs on a per-pixel basis. The likelihood of observing a lesion at any particular location is in fact very low (see Figure \ref{fig:lesion_prob} (b)) and more or less uniform. It is this uniformity that removes the bias towards any particular location. Results comparing a uResNet with out center shifting sampling are shown in Table \ref{tab:uResNet_comparison}.

\begin{table}[!t]
\begin{center}
{\footnotesize
\renewcommand{\tabcolsep}{5pt}
\begin{tabular}{ |l|ccc|c| }
\hline \hline
					&\textbf{uResNet}	&\textbf{uResNet2}	&\textbf{uResNet\_NoC}	&\textbf{Expert}\\\hline \hline
WMH Dice (std)		&{69.5\textbf(16.1)}			&\textbf{69.6(16.1)}&66.9(18.1)				&-				\\ 
Stroke Dice (std)	&40.0(25.2)			&\textbf{40.2}(27.7)&28.9\textbf{(22.3)}				&-				\\ 
WMH $R^2$			&\textbf{0.951}		&\textbf{0.951}		&0.948					&-				\\ 
Stroke $R^2$		&\textbf{0.791}		&0.761				&0.710					&-				\\ 
WMH Trend			&\textbf{0.89x+0.07}&\textbf{0.89x}-0.08			&\textbf{0.89x}+0.15				&-				\\ 
Stroke Trend		&\textbf{0.58x+0.01}&0.55x-\textbf{0.01}&0.52x+0.07				&-				\\ 
CC D-Fazekas 		&0.770				&\textbf{0.776}		&0.771					&0.774			\\ 
CC PV-Fazekas		&0.778				&\textbf{0.783}		&0.777					&0.765			\\ 
CC Fazekas			&0.824				&\textbf{0.831}		&0.823					&0.819			\\ 
CC MMSE				&0.364				&0.373				&0.366					&0.372			\\ 
\hline
\end{tabular}}
\caption{Mean Dice scores of WMH and stroke (standard deviation in parenthesis), correlation analysis between expert and automatic volumes ($R^2$ and trend), and correlation with clinical variables. No statistical significance between uResNet and uResNet2 was observed ($p>0.05$), while there was a statistically significant difference ($p<0.001$) between patch off-center sampling (uResNet) and regular no off-center sampling (uResNet\_NoC).}
  \label{tab:uResNet_comparison}
\end{center}
\end{table}

\section{Dice results for different inputs}\label{sec:app}

Both CNN approaches, uResNet and DeepMedic, can easily be trained using one or several inputs. Table \ref{tab:dice_uRes_DM} provides Dice overlap results of using different input channels in both CNN approaches are provided. As it can be appreciated DeepMedic can narrow the Dice overlap gap with uResNet if several inputs are provided. However, as discussed before, obtaining and generating these extra inputs limits clinical applicability and also adds additional computational costs to the whole segmentation framework.

\begin{table}[ht]
\begin{center}
{\footnotesize
\renewcommand{\tabcolsep}{5pt}
\begin{tabular}{ |l|ccc|ccc| }
\hline \hline
\textbf{Input}		&\multicolumn{3}{|c|}{\textbf{WMH} }	&\multicolumn{3}{|c|}{\textbf{Stroke} }	\\
\textbf{channels}&\textbf{uResNet}&\textbf{DeepMedic}&\textbf{Diff}	&\textbf{uResNet}&\textbf{DeepMedic}&\textbf{Diff}\\\hline \hline
F					&\textbf{69.5}		&66.3	&\textit{3.2}			&\textbf{40.0}		&31.1			&\textit{8.9}	\\ 
F-T1				&\textbf{69.7}		&67.6	&\textit{2.1}			&\textbf{35.5}		&34.3			&\textit{1.2}	\\ 
F-CS 				&\textbf{69.1}		&66.6	&\textit{2.5}			&\textbf{36.7}		&35.1			&\textit{1.6}	\\ 
F-WM				&\textbf{69.4}		&68.2	&\textit{1.2}			&33.0				&\textbf{35.9} &\textit{-2.9}	\\ 
F-CS-WM 			&\textbf{69.3}		&68.0	&\textit{1.3}			&\textbf{38.4}		&37.8			&\textit{0.6}	\\ 
F-T1-CS-WM			&\textbf{69.6}		&68.4	&\textit{1.2}			&\textbf{40.2}		&36.0			&\textit{4.2}	\\ 
\hline
\end{tabular}}
\caption{Mean Dice scores of WMH and stroke, for different inputs with uResNet and DeepMedic. Difference in Dice score between the two methods is given in italics. F: FLAIR image, CS: cerebro-spinal track atlas, WM: white matter probability map, T1: T1 weighted image.}
  \label{tab:dice_uRes_DM}
\end{center}
\end{table}

\section{Additional DeepMedic experiments}\label{sec:appC}
DeepMedic experiments that aim to approximate the sampling scheme used by uResNet were carried out, where several sampling weights were tested for DeepMedic.  
A direct comparison of per-class patch sampling is not straight forward between the proposed method and DeepMedic, and furthermore it can be misleading. For instance, in the work proposed here a sampling rate of 80-20\% of WHM-stroke patches is used, each patch has a size of 64 by 64 voxels and uResNet makes a prediction of a 64 by 64 patch of the label space during training (it is fully convolutional and uses padded convolutions throughout). This means that each patch used in uResNet has a label map that due to its size inevitably contains a large amount of healthy tissue. Therefore we do not sample specifically from healthy regions. On the other hand, DeepMedic trains with segments that have a label space of 9 by 9 by 9 voxels, therefore it is far less likely that healthy tissue is included in non-healthy samples and thus healthy segments need to be sampled. Nonetheless, different per-class sampling rates, as well as other hyper-parameter settings with DeepMedic were explored.

Some of DeepMedic's default hyper-parameter values are: learning rate of 1e-3, RmsProp optimizer, sampling form of foreground/background (diseased/healthy tissue) and sampling rate of {[0.5, 0.5]} (healthy and diseased tissue). The different sampling rates tested with DeepMedic in our experiments to approximate uResNet setup were {[0.5, 0.1, 0.4]}, {[0.5, 0.25, 0.25]}, {[0.33, 0.13, 0.53]} and {[0.33, 0.33, 0.3]}, for healthy, WMH and stroke tissue, respectively. Additionally, learning rate values explored were in the range of 1.9e-2 to 1e-4, with RMSprop, Adam or SGD as optimizer. Changing the sampling rates from the default generally produced unstable results, with either failing to converge or producing poorer overlap values than with the default settings.
In total, 14 different additional DeepMedic train/test runs were performed,
out of which only two converged, both using a sampling rate of {[0.33, 0.33,0.33]}. Dice overlap results by these experiments were of 60.7 and 29.9, for WMH and stroke, respectively, in one instance and 59.4 and 29.5 in the other.
These unstable results might be due to the tuning of additional meta parameters, such as the optimizer, learning rate or regularization.
Therefore, presented DeepMedic results were obtained with default hyper-parameters, which were the best results obtained in our experiments.

\bibliographystyle{abbrv}
\bibliography{Bibliography}

\end{document}